\documentclass{article}

     \PassOptionsToPackage{numbers, sort&compress}{natbib}

 \usepackage[dblblindworkshop, final]{neurips_2025}

\usepackage[utf8]{inputenc} 
\usepackage[T1]{fontenc}    
\usepackage{hyperref}       
\usepackage{url}            
\usepackage{booktabs}       
\usepackage{amsfonts}       
\usepackage{nicefrac}       
\usepackage{microtype}      
\usepackage{xcolor}         

\usepackage{amsmath, amssymb}
\usepackage{graphicx}
\usepackage{caption}
\usepackage{subcaption}
\usepackage{enumitem}

\title{Compressing Biology: Evaluating the Stable Diffusion VAE for Phenotypic Drug Discovery}

\author{%
  Télio Cropsal \\
  Department of Computer Science and Engineering\\
  Chalmers University of Technology and University of Gothenburg\\
  Gothenburg, SE \\
  \texttt{telio@chalmers.se} \\
  \And
  Rocío Mercado \\
  Department of Computer Science and Engineering\\
  Chalmers University of Technology and University of Gothenburg\\
  Gothenburg, SE \\
  \texttt{rocio.mercado@chalmers.se} \\
}

\begin{document}

\maketitle

\begin{abstract}
  High-throughput phenotypic screens generate vast microscopy image datasets that push the limits of generative models due to their large dimensionality. Despite the growing popularity of general-purpose models trained on natural images for microscopy data analysis, their suitability in this domain has not been quantitatively demonstrated. We present the first systematic evaluation of Stable Diffusion's variational autoencoder (SD-VAE) for reconstructing Cell Painting images, assessing performance across a large dataset with diverse molecular perturbations and cell types. We find that SD-VAE reconstructions preserve phenotypic signals with minimal loss, supporting its use in microscopy workflows. To benchmark reconstruction quality, we compare pixel-level, embedding-based, latent-space, and retrieval-based metrics for a biologically informed evaluation. We show that general-purpose feature extractors like InceptionV3 match or surpass publicly available bespoke models in retrieval tasks, simplifying future pipelines. Our findings offer practical guidelines for evaluating generative models on microscopy data and support the use of off-the-shelf models in phenotypic drug discovery.
\end{abstract}

\section{Introduction}

Phenotypic drug discovery is a strategy in drug development that identifies drug candidates by directly observing their effects in biological systems without requiring prior knowledge of molecular targets \cite{vincent2022phenotypic}. By focusing on measurable phenotypic changes induced by molecular perturbations, this approach has historically led to the discovery of several clinically relevant drugs, such as the anti-malarial artemisinin \cite{miller2011artemisinin}. Recent advances in high-throughput microscopy, e.g., Cell Painting \cite{bray2016cell}, have accelerated phenotypic screening pipelines \cite{rietdijk2021phenomics}. In Cell Painting, cells are stained with multiple fluorescent dyes that mark distinct subcellular components, enabling the capture of rich morphological profiles via automated, low-cost fluorescence microscopy. A single lab can thus yield millions of high-res, multi-channel images under diverse conditions. These images can be processed with tools such as CellProfiler \cite{stirling2021cellprofiler} to extract thousands of morphological features per condition. However, the sheer dimensionality of the resulting profiles pose significant analytical challenges, motivating the development of methods that can uncover subtle phenotypic patterns at scale.

Deep learning (DL) methods have become popular tools for addressing these challenges, initially via representation learning methods to extract meaningful features from raw images \cite{Hofmarcher2019, Nguyen2024, sanchez2023cloome}. More recently, the success of generative models has inspired efforts to simulate Cell Painting images, with the aim of reducing experimental cost and enabling virtual screening \cite{cross2023class}. Several studies have shown promising results in generating realistic microscopy images conditioned on molecular \cite{Yang2021} or genetic \cite{navidi2024morphodiff} perturbations. Despite this progress, generative modeling of microscopy data remains computationally challenging due to its high dimensionality. Direct pixel-space generation is costly, leading many to restrict image generation tasks to small crops centered around individual nuclei \cite{palma2025predicting, bourou2024phendiff, zhang2025cellfluxsimulatingcellularmorphology} or to adopt latent diffusion approaches \cite{navidi2024morphodiff, papanastasiou2024confounder}. Latent diffusion models, exemplified by Stable Diffusion (SD) \cite{rombach2022highresolutionimagesynthesislatent}, tackle complexity by first mapping images to a compressed latent representation using a variational autoencoder (VAE) \cite{Kingma2019}, then performing diffusion-based generative modeling within this lower-dimensional space. At inference, images are generated in latent space and decoded back to image space, with the VAE acting as a bottleneck for the final reconstruction quality.

Reliable generation of microscopy images is critical for downstream biological interpretation, and the SD-VAE is increasingly used for this purpose \cite{navidi2024morphodiff, papanastasiou2024confounder}. However, the reconstruction quality of SD-VAE-generated images has not been systematically evaluated. This raises concerns about whether meaningful biological information may be lost during the encoding-decoding process, especially when applied to out-of-distribution microscopy data such as Cell Painting images. To address this gap, this work evaluates the reconstruction fidelity of SD-VAE on Cell Painting images using a recently established benchmark.

Our main contributions address this gap and are as follows:

\begin{itemize}[left=0pt, itemsep=-1pt, topsep=-1pt]
    \item \textbf{First systematic evaluation of SD-VAE on microscopy data:} We evaluated SD-VAE, trained primarily on natural images, on $>$1M Cell Painting crops spanning two cell types (A549 and U2OS) and diverse perturbations. We demonstrate that SD-VAE reconstructions retain phenotypic signals with minimal degradation, validating its use in microscopy image generation.
    
    \item \textbf{A general evaluation framework for generative models of microscopy images:} We systematically compare pixel-level metrics, e.g., mean absolute error (MAE) and earth mover's distance (EMD), with feature-space and latent-space metrics, e.g., Kullback-Leibler divergence (KLD) and Fréchet Inception distance (FID), as well as retrieval-based evaluations, presenting a robust framework that researchers can model future validation studies on. 

    \item \textbf{General-purpose feature extractors rival domain-specific models:} We find that InceptionV3 embeddings match or exceed those from the publicly-available microscopy-specialized OpenPhenom model 
    on biologically relevant retrieval tasks, suggesting that general-purpose feature extractors may be sufficient to capture subtle phenotypic variations.
    
\end{itemize}

\section{Method}

Our benchmarking pipeline uses multiple quantitative metrics to evaluate the effectiveness of SD-VAE in compressing and reconstructing microscopy data. 
We use two distinct pre-trained models for image featurization: InceptionV3~\cite{szegedy2015rethinkinginceptionarchitecturecomputer}, trained on natural images, and OpenPhenom~\cite{Kraus2024MaskedAF, Recursion2024OpenPhenom}, trained specifically on Cell Painting images. We do not compare the feature extractors with CellProfiler \cite{stirling2021cellprofiler} due to its high computational cost and the complexity of integrating it into scalable, automated GPU-based workflows. Instead, we rely on deep learning-based models, which are better suited to our infrastructure and more representative of how generative models are typically deployed in production settings. We assess reconstruction quality as described in Section \ref{sec:evaluation}. 

\begin{figure}
  \centering
  \includegraphics[width=0.95\textwidth, trim={0 7 0 7}, clip]{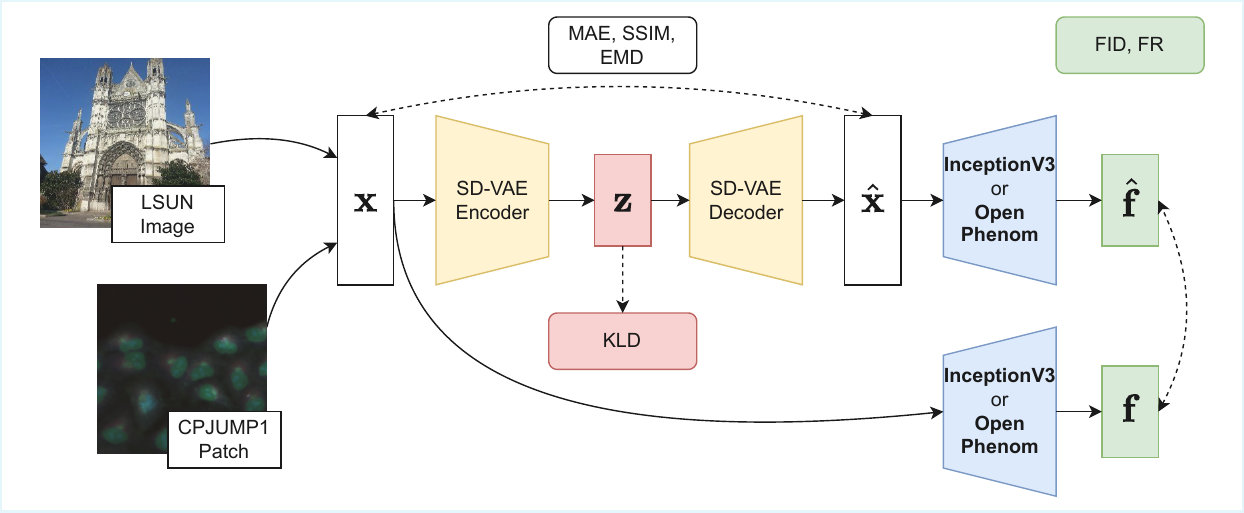} 
  \vspace{-5pt}
  \caption{Overview of our evaluation pipeline. Input images, including Cell Painting and natural images, are encoded into the latent space using the Stable Diffusion VAE (SD-VAE). These latent representations are decoded to reconstruct the images. Original and reconstructed images are compared using channel-wise metrics: mean absolute error (MAE), structural similarity index measure (SSIM), and Earth mover's distance (EMD). Both sets of real and reconstructed images are passed through InceptionV3 and OpenPhenom to extract feature embeddings, which are used to compute the Fréchet Inception distance (FID) and fraction retrieved (FR) of perturbations against negative controls. Latent vectors are further used to compute the Kullback-Leibler divergence (KLD).}
\end{figure}

\subsection{Pre-trained models}

\paragraph{Stable Diffusion VAE}

We evaluate the VAE associated with the model \textit{stable-diffusion-v1-4}~\cite{CompVis2022StableDiffusion, Rombach_2022_CVPR}, 
which was trained on natural images of varying dimensions (256x256 and 512x512) from the \textit{laion2B-en} dataset. This autoencoder employs a relative downsampling factor of 8. For example, an RGB image with a resolution of (3x256x256) would be mapped to a latent tensor of shape (4x32x32). Throughout all experiments the VAE weights are kept frozen and used only for encoding and decoding.

\paragraph{InceptionV3} As a baseline feature extraction method we use the \textit{torchvision} implementation of InceptionV3 \cite{szegedy2015rethinkinginceptionarchitecturecomputer}, which has been pretrained on the ImageNet dataset \cite{deng2009imagenet}. To evaluate the capabilities of models for image generation, the Fréchet distance is typically applied to features extracted from the layer just before the final classification layer of an InceptionV3 network. The empirical Fréchet distance computed between the means and covariances of these features following real and generated distributions is known as the Fréchet Inception distance (FID) \cite{heusel2018ganstrainedtimescaleupdate} (see Appendix \ref{sec:metrics}).

\paragraph{OpenPhenom}
As an alternative to CellProfiler, we utilize a pre-trained channel-agnostic masked autoencoder (CA-MAE) called OpenPhenom, which employs a ViT-S/16 encoder backbone to generate embeddings for Cell Painting images \cite{Kraus2024MaskedAF}. OpenPhenom is a fully open-access model available on HuggingFace~\cite{Recursion2024OpenPhenom}, 
trained on $>$3M 256×256 image crops from publicly available Cell Painting datasets with genetic perturbations, including RxRx3 (HUVEC cell line) \cite{Fay2023RxRx3PM} and the JUMP-CP CRISPR and ORF subsets (U2OS cell line) \cite{chandrasekaran2023jump}. The model was trained for 100 epochs using a CA-MAE architecture with a 25M parameter ViT-S/16 encoder and six dedicated decoders (one per channel). The training channels include \textit{mitochondria}; \textit{DNA} (nucleus); \textit{RNA}; \textit{endoplasmic reticulum (ER)}; \textit{actin, Golgi and plasma membrane (AGP)}; and a non-fluorescent Brightfield channel. During training, each 16×16 patch from each channel is tokenized independently, resulting in 1,536 tokens for a 6-channel image, with 384 unmasked tokens visible under a 75\% masking ratio. OpenPhenom can generate embeddings either for the entire image or per individual channel.

\subsection{Data}

We evaluate the performance of the SD-VAE on one out-of-distribution Cell Painting dataset and, as a control, one in-distribution natural image dataset. For data preprocessing details, see Appendix \ref{sec:preprocessing}.

\paragraph{CPJUMP1} 

This dataset \cite{Chandrasekaran2022ThreeMI} includes replicated plates containing both chemical and genetic perturbations with known mechanisms of action or molecular targets. These replicates are tested under different experimental conditions, including two cell lines (A549 and U2OS) and two exposure durations, specifically 24 hours and 48 hours for compound treatments, representing short and long time points. For this study, we used a subset of the dataset that includes only chemical perturbations with 100\% cell seeding and parental cell lines (66,048 center-cropped 1024x1024 images among 16 plates exposed to 307 unique perturbations, including DMSO controls). Each image contains 5 channels, corresponding in this dataset to fluorescent stains of the mitochondria (\textit{Mito}), actin, Golgi, and plasma membrane (\textit{AGP}), nucleoli and cytoplasmic RNA (\textit{RNA}), endoplasmic reticulum (\textit{ER}), and nuclear DNA (\textit{DNA}).  

\paragraph{LSUN} As a control, we also assess the performance of SD-VAE on high-res natural images which closely resemble the VAE's training data. Specifically, we utilize two subsets from the LSUN \cite{yu2016lsunconstructionlargescaleimage} dataset: the classroom subset, comprising 166,419 images, and the outdoor church subset, containing 126,200 images. All images are already resized so that the smaller dimension is 256 pixels. Images consist of the three standard RGB channels. Our evaluation on this dataset uses the same set of metrics for consistency.


\subsection{Evaluation}
\label{sec:evaluation}

Images or patches with a 256×256 pixel resolution are first passed through the encoder and decoder of the SD-VAE. The resulting reconstructions are then compared to the original images to assess reconstruction quality. Moreover, both the original and reconstructed images are processed through two pre-trained networks, InceptionV3 and OpenPhenom, for independent feature extraction to assess the reconstruction of morphological profiles via either scheme. Details and equations for all metrics are provided in Appendix \ref{sec:metrics}.

\paragraph{Channel- and distribution-wise}

We compute the mean absolute error (MAE), structural similarity index measure (SSIM), and Earth mover’s distance (EMD) between each channel of the real and reconstructed images. While MAE provides a straightforward pixel-wise error, SSIM captures perceptual differences by considering structural information, and EMD evaluates how closely the distributions of pixel intensities match between the original and reconstructed images. Conversely, the Fréchet Inception distance \cite{heusel2018ganstrainedtimescaleupdate} is a widely used distribution-based metric for evaluating image generation models. It measures the difference between the distributions of real and generated images by comparing the means and covariances of features extracted from the Inception network.

\paragraph{Regularized latent space} To complement the aforementioned metrics and evaluate the quality of the learned latent space by the SD-VAE, we evaluate the Kullback-Leibler divergence (KLD) between the samples in the latent space and a standard multivariate Gaussian distribution to assess how well the latent space is regularized, indicating how easily it can be learned, e.g., by a diffusion model.

\paragraph{Information retrieval}

To better reflect the practical application of generative models in phenotypic drug discovery, we follow the evaluation procedure described by \citet{Chandrasekaran2022ThreeMI} and \citet{Kalinin2024AVI}. This approach evaluates the quality of learned embeddings through a retrieval task that identifies replicates of the same perturbation, target, or mechanism of action (MoA). Here we compare the InceptionV3 network and the pre-trained masked autoencoder OpenPhenom. The retrieval task is performed separately on embeddings obtained from real and reconstructed images. Plate-specific batch effects are accounted for and mitigated through post-processing of the features extracted from InceptionV3 or OpenPhenom using negative control wells, as detailed in Appendix \ref{sec:postprocessing}. Our evaluation focuses solely on phenotypic activity, as it serves as a challenging sanity check for downstream applications. Specifically, we assess whether replicate profiles for a given perturbation can be distinguished from replicate profiles under control conditions, using negative controls as the reference. Hence, the fraction retrieved (FR) metric quantifies the proportion of perturbations that can be reliably distinguished from negative controls. For this purpose, we employ the \textit{copairs} library \cite{Kalinin2024AVI}, specifically designed for retrieval-based analysis.
Features extracted from each pre-trained network are aggregated across imaging sites to produce a single feature vector per well. Features extracted from InceptionV3 are concatenated across the two distinct 3-channel input images, resulting in a per-sample dimensionality of $2 \times 2048 = 4096$. Similarly, OpenPhenom features are concatenated along the channel axis, yielding a per-sample dimensionality of $5 \times 384 = 1920$. See Appendix \ref{sec:postprocessing} for key post-processing details. 

\section{Results}

\subsection{SD-VAE reconstructs images well in terms of standard reconstruction metrics}

Cell Painting images show a lower MAE between original and reconstructed samples (Figure \ref{fig:mae_emd_combined}), indicating that SD-VAE effectively reconstructs them. This aligns with expectations, as Cell Painting images typically contain simpler and more structured visual patterns than natural images, with certain channels often displaying a relatively consistent background. Supporting metrics such as EMD, FID, and SSIM also show similar value ranges across both image types (Figures \ref{fig:mae_emd_combined}--\ref{fig:ssim}), reinforcing that reconstructed Cell Painting images are visually faithful to the originals.

\subsection{Biological signal preserved after SD-VAE reconstruction}

While metrics like MAE confirm better pixel-level reconstruction of cell images than natural images, they do not capture the reconstruction of relevant biological features. Instead, metrics like the fraction retrieved (FR) can help us infer if biological signal is preserved following application of the SD-VAE. Notably, when evaluating the FR across different cell lines and time points (Table \ref{table:aggregated}), we found that the reconstructed images from SD-VAE did not lead to a significant drop in FR, even demonstrating a slight increase in many cases. This suggests that, despite the reconstruction process, the biological signal remains sufficiently intact to distinguish between negative controls and perturbations.

\begin{table}[ht]
\centering
\caption{Fraction retrieved (FR) across cell types and time points, with per-well median aggregation and plate-level mean scaling of DL features. 
The reported values are averaged over three independent runs of the complete pipeline (SD-VAE and information retrieval). The largest std. dev. is $\sigma=0.008$. Experiments with original images by design show no variation ($\sigma=0$). The best value for each cell line and time point is shown in bold. The bottom row compares performance on the same task for a traditional feature extraction method.}
\medskip
\begin{tabular}{ll|cccc}
\toprule
 & & \multicolumn{2}{c}{\textbf{A549}} 
& \multicolumn{2}{c}{\textbf{U2OS}} \\
\cmidrule(lr){3-4} \cmidrule(lr){5-6}
\textbf{Features} & \textbf{Data} & \textbf{24h} & \textbf{48h} & \textbf{24h} & \textbf{48h} \\
\midrule
OpenPhenom & Original       & 0.722 & 0.882 & 0.817 & 0.660 \\
OpenPhenom & SD-VAE         & 0.729 & 0.879 & 0.836 & 0.697 \\
InceptionV3 & Original      & 0.873 & \textbf{0.961} & 0.837 & \textbf{0.837} \\
InceptionV3 & SD-VAE        & \textbf{0.906} & 0.951 & \textbf{0.847} & \textbf{0.837} \\
\midrule
CellProfiler \cite{Chandrasekaran2022ThreeMI} & Original  & 0.761 & 0.954 & 0.775 & 0.663 \\
\bottomrule
\end{tabular}
\label{table:aggregated}
\end{table}

\subsection{KLD suggests microscopy latents are less regularized than those of natural images}

We observe that Cell Painting images exhibit a higher Kullback–Leibler divergence (KLD) between their latent representations and an isotropic Gaussian prior, compared to natural images (Figure \ref{fig:fid_kl_combined}). This indicates that the latent space for microscopy data is less regularized and the encoded representations deviate more from the prior distribution. In contrast, natural images produce latent vectors that are closer to the prior. This difference suggests that it is more difficult for the model to compress the complex, biologically rich content of Cell Painting images into a smooth, well-structured latent space. As a result, tasks that rely on latent representations may be affected by this reduced regularization, e.g., this may complicate the training of downstream latent diffusion models.

\section{Discussion}

While metrics like MAE, SSIM, EMD, and FID are useful for assessing low-level similarity, they provide limited insight into the preservation of biological signal. This motivates the use of more biologically grounded and interpretable evaluation strategies, such as measuring the FR of perturbations against negative controls, to assess whether reconstructions retain relevant phenotypic information.

Our experiments show that general-purpose feature extractors, such as InceptionV3, can perform on par with, and in some cases better than, domain-specific models like OpenPhenom in tailored retrieval tasks. This suggests that models pre-trained on natural images may be sufficiently effective for evaluating generative models in the context of phenotypic drug discovery, reducing the need for specialized feature extractors. The relatively weaker performance of OpenPhenom relative to InceptionV3 can be attributed to several factors. First, there is a discrepancy between the training and inference conditions: during inference, only five channels are provided while the Brightfield channel (used during OpenPhenom training) is omitted. Second, the OpenPhenom model used here is a comparatively small model trained on a limited dataset, unlike the other larger CA-MAE variants which are unfortunately not publicly available. Overall, our results support the use of FID as a practical and reliable metric during model development and evaluation, as evidenced by its alignment with the FR and the demonstrated effectiveness of Inception features in the retrieval task.

Note that in this study, we deliberately avoid evaluating the SD-VAE latent space; this is because evaluating the latent space in greater depth beyond KLD would require specialized methods. In this work, we instead rely on existing models (InceptionV3 and OpenPhenom) to evaluate the reconstructed images rather than the latent space. This setup better reflects future use cases of generative models, where new samples are generated and assessed by surrogate models in an automated fashion. Additionally, it has been shown that the latent space of SD-VAE is not strongly semantically regularized; rather, it serves as a compressed representation of the original image, removing redundant information while preserving spatial structure \cite{kouzelis2025eqvaeequivarianceregularizedlatent}. We leave for future work a comparison between SD-VAE and other dimensionality reduction techniques, since it would also be necessary to demonstrate that the resulting latent space is suitable for generating Cell Painting images. This has already been shown with MorphoDiff \cite{navidi2024morphodiff}, although their evaluation did not isolate the performance of SD-VAE. For similar reasons, we have not fine-tuned SD-VAE, as the straightforward approach is known to be ineffective \cite{leng2025repaeunlockingvaeendtoend}.

There are a few additional limitations to our approach. First, our dataset contains five distinct channels, but both the VAE and the InceptionV3 model require 3-channel inputs. To address this, we duplicated one of the channels to create two separate 3-channel combinations, allowing us to use all five channels while maintaining compatibility with the models. Although effective, this approach is somewhat arbitrary and may not be optimal. Future work could investigate more systematic strategies for channel grouping or selection. Second, there is the risk of data leakage. Using negative controls in the post-processing pipeline is standard practice for limiting batch effects. Since FR involves distinguishing perturbations from negative controls, this may be a form of data leakage. Moreover, we observed an increase in the FR metric across most experiments involving reconstructed images. This trend may suggest a denoising effect, as the reconstruction process could be removing noise or artifacts in ways that improve FR performance. While we acknowledge this potential limitation, addressing it in depth is also left for future work.

\section{Conclusion}

When working with high-dimensional Cell Painting images, SD-VAE appears to be a promising approach for image generation that mostly preserves biological signal while reducing the dimensionality of the data. 
This is critical because the generated images or even latents are typically be used as input for downstream models that aim to identify meaningful patterns in this high-dimensional data. With this work, we provide a framework to ensure that the VAE, or any other generative model, can be adequately integrated into the overall workflow without significant degradation of biological signal. Our work further supports the use of SD-VAE and general metrics like FID in Cell Painting analysis workflows without the need for specialized training of bespoke models.

\section*{Acknowledgments} 
TC and RM acknowledge the funding provided by the Wallenberg AI, Autonomous Systems, and Software Program (WASP), supported by the Knut and Alice Wallenberg Foundation. The computations and data storage were enabled by resources provided by Chalmers e-Commons at Chalmers.
The computations and data storage were enabled by resources provided by the National Academic Infrastructure for Supercomputing in Sweden (NAISS), partially funded by the Swedish Research Council through grant agreement no. 2022-06725.

\section*{Code and Data Availability}
For implementation details and source code, please refer to our anonymized GitHub repository: 
\href{https://github.com/cctelio/compressing-biology}{\texttt{compressing-biology}}. Pre-trained models and datasets used can be downloaded from:

\begin{itemize}
    \item SD-VAE: \href{https://huggingface.co/CompVis/stable-diffusion-v1-4}{huggingface.co/CompVis/stable-diffusion-v1-4}
    \item InceptionV3: \href{https://docs.pytorch.org/vision/main/models/inception.html}{docs.pytorch.org/vision/main/models/inception.html}
    \item OpenPhenom: \href{https://huggingface.co/recursionpharma/OpenPhenom}{huggingface.co/recursionpharma/OpenPhenom}
    \item CPJUMP1: \href{https://cellpainting-gallery.s3.amazonaws.com/index.html\#cpg0000-jump-pilot/source_4/images/2020_11_04_CPJUMP1/}{\texttt{cellpainting-gallery.s3.amazonaws.com}}
    \item LSUN: \href{https://docs.pytorch.org/vision/main/generated/torchvision.datasets.LSUN.html}{docs.pytorch.org/vision/main/generated/torchvision.datasets.LSUN.html}
\end{itemize}

{
\small
\bibliographystyle{plainnat}
\bibliography{main}
}


\appendix

\newpage
\section{Technical Appendices and Supplementary Material}
\subsection{Notation}

Let \( \mathbf{x} \in \mathbb{R}^{C \times H \times W} \) denote the ground truth image and \( \hat{\mathbf{x}} \in \mathbb{R}^{C \times H \times W} \) the reconstructed image. The following notation is used:

\begin{itemize}
    \item \( C \): number of channels
    \item \( H, W \): image height and width
    \item \( N = H \times W \): number of pixels per channel
    \item \( \mu_{\mathbf{x}}, \mu_{\hat{\mathbf{x}}} \): local means
    \item \( \sigma_{\mathbf{x}}^2, \sigma_{\hat{\mathbf{x}}}^2 \): local variances
    \item \( \sigma_{\mathbf{x}\hat{\mathbf{x}}} \): local covariance
    \item \( C_L \): constant to stabilize luminance comparison
    \item \( C_C \): constant to stabilize contrast and structure comparison
    \item \( \mu_i, \log \sigma_i^2 \): mean and log-variance of latent variable \( z_i \)
    \item \( d \): dimensionality of the latent space
    \item \( \boldsymbol{\mu}_r, \boldsymbol{\Sigma}_r \): mean and covariance of real image features
    \item \( \boldsymbol{\mu}_g, \boldsymbol{\Sigma}_g \): mean and covariance of generated image features
\end{itemize}

\subsection{Evaluation Metrics}
\label{sec:metrics}

Below we define the pixel- and distribution-based metrics used in this study:

\begin{enumerate}
\item \textbf{Mean absolute error (MAE)}  
The average absolute difference between corresponding pixels:
\begin{equation}
\text{MAE}_c = \frac{1}{HW} \sum_{i=1}^{H} \sum_{j=1}^{W} \left| \mathbf{x}_{c, i, j} - \hat{\mathbf{x}}_{c, i, j} \right|
\end{equation}

\item \textbf{Structural similarity index (SSIM)}  
A perceptual similarity measure combining luminance, contrast, and structure:
\begin{equation}
\text{SSIM}_c(\mathbf{x}, \hat{\mathbf{x}}) = \frac{(2\mu_{\mathbf{x}} \mu_{\hat{\mathbf{x}}} + C_L)(2\sigma_{\mathbf{x}\hat{\mathbf{x}}} + C_C)}{(\mu_{\mathbf{x}}^2 + \mu_{\hat{\mathbf{x}}}^2 + C_L)(\sigma_{\mathbf{x}}^2 + \sigma_{\hat{\mathbf{x}}}^2 + C_C)}
\end{equation}

\item \textbf{Earth mover's distance (EMD)}  
The average absolute difference between sorted pixel intensities:
\begin{equation}
\text{EMD}_c = \frac{1}{N} \sum_{k=1}^{N} \left| \text{sort}(\mathbf{x}_c)_k - \text{sort}(\hat{\mathbf{x}}_c)_k \right|
\end{equation}

\item \textbf{Kullback–Leibler divergence (KLD)}  
The divergence between the latent distribution and a standard Gaussian:
\begin{equation}
\text{KL} = \frac{1}{2} \sum_{i=1}^{d} \left( \mu_i^2 + \exp(\log \sigma_i^2) - \log \sigma_i^2 - 1 \right)
\end{equation}

\item \textbf{Fréchet inception distance (FID)}  
The distance between real and generated feature distributions:
\begin{equation}
\text{FID} = \left\| \boldsymbol{\mu}_r - \boldsymbol{\mu}_g \right\|^2 + \text{Tr}\left( \boldsymbol{\Sigma}_r + \boldsymbol{\Sigma}_g - 2\left( \boldsymbol{\Sigma}_r \boldsymbol{\Sigma}_g \right)^{1/2} \right)
\end{equation}
\end{enumerate}

\subsection{Data Pre-Processing Details}
\label{sec:preprocessing}

Images from the CPJUMP1 dataset are processed similar to RxRx3-Core \cite{kraus2025rxrx3corebenchmarkingdrugtargetinteractions}. The provided illumination correction arrays are applied, and images are saved as \textit{uint8} and compressed as PNGs. LSUN images are already optimized for DL pipelines.
Images are standardized to 256×256 pixels to ensure consistent resolution: LSUN images are directly resized to the target resolution via interpolation if needed, whereas CPJUMP1 images are divided into 256×256 pixel patches.
As SD-VAE and InceptionV3 expect the standard 3-channel (RGB) inputs, the 5-channel Cell Painting images are handled differently: one of the five channels (RNA, selected randomly) is duplicated to create 6 channels. These are then split into two separate 3-channel images 
following the approach of \citet{papanastasiou2024confounder}. For the OpenPhenom model, images are pre-processed using self-standardization, as recommended by \citet{Kraus2024MaskedAF}, and kept in the 5-channel format as the model is channel agnostic.

\subsection{Data Post-Processing Details}
\label{sec:postprocessing}
Extracted feature vectors are post-processed using a pipeline inspired by the typical variation normalization (TVN) method \cite{Ando2017ImprovingPM} recommended by OpenPhenom. Post-processing is done as follows:

\begin{itemize}
    \item Fit a sequence of preprocessing steps, including scaling, principal component analysis (PCA), and variance thresholding, to all negative control samples, using the highest feasible dimensionality for PCA.
    \item Apply the fitted sequence of steps to all samples
    \item Scale all features within each plate using the negative controls from the same plate.
\end{itemize}

We did not use the post-processing steps recommended for the CPJUMP1 dataset, as it is specifically tailored to CellProfiler-derived features. Instead, we adapted the OpenPhenom recommended post-processing steps to account for the limited number of negative controls in our dataset, which restricts the maximum dimensionality of the PCA step.

Note that while the phenotypic activity benchmark offers an interpretable and practical framework for evaluating DL models in phenotypic drug discovery, performance is highly sensitive to the design of the post-processing pipeline. For example, pipelines optimized for CellProfiler features may not generalize well to DL-derived features, and the effectiveness of each step often depends on the availability and quality of metadata. Even seemingly minor choices, such as aggregating features using the mean versus the median (Table \ref{table:aggregated_full}), can shift the relative performance of models. These findings highlight that pipeline components should not be treated as modular or interchangeable. Each step must be carefully designed in the context of the full workflow, especially when generative models are involved.

\subsection{Additional Results}
\label{sec:additional_results}

\begin{figure}[h!]
  \centering
  \begin{subfigure}[t]{0.49\textwidth}
    \includegraphics[width=\textwidth]{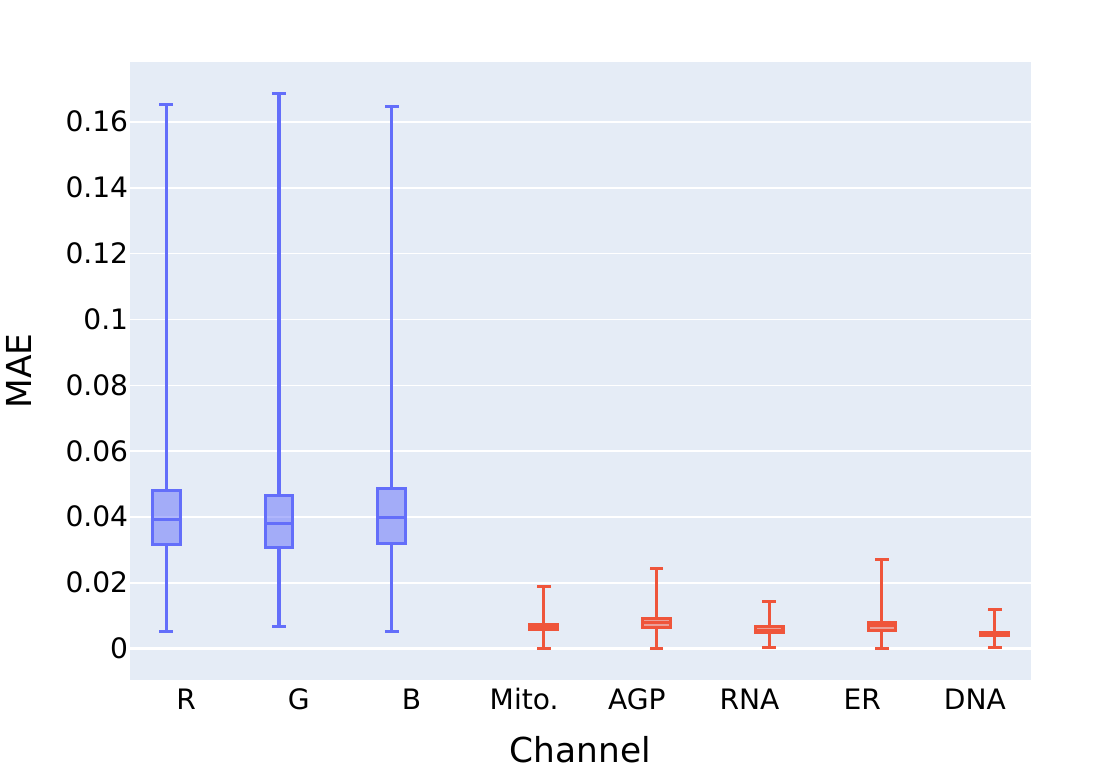}
  \end{subfigure}
  \hfill
  \begin{subfigure}[t]{0.49\textwidth}
    \includegraphics[width=\textwidth]{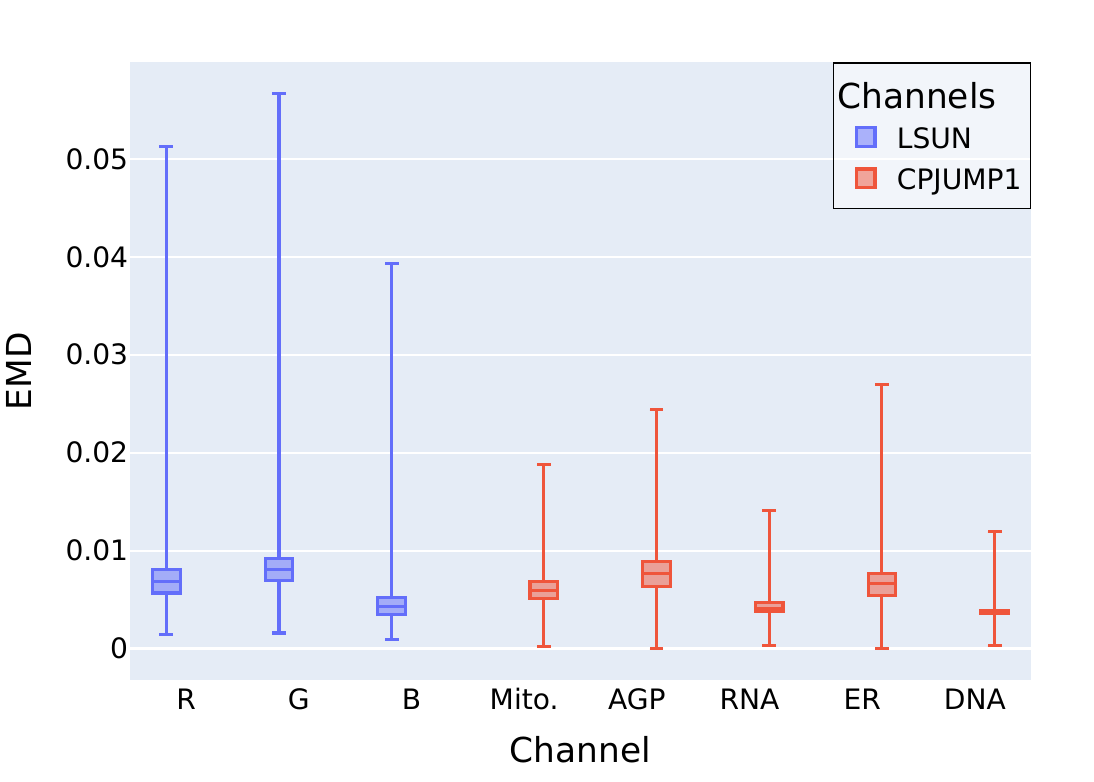}
  \end{subfigure}
  \vspace{-3pt}
  \caption{Box plots showing MAE (left) and EMD (right) values across the LSUN and CPJUMP1 datasets and their channels, computed after a single run of SD-VAE applied to all images. The central line within each box shows the median, while the box boundaries represent the 1st and 3rd quartiles.}
  \label{fig:mae_emd_combined}
\end{figure}

In Figure \ref{fig:mae_emd_combined} we show the results for MAE and EMD across the LSUN and CPJUMP1 datasets, illustrating how these metrics are similar in range for both types of datasets. LSUN images were used to establish a baseline for these metrics, since the Church and Classroom subsets we looked at are natural images similar to the images used to train the SD-VAE. 

\begin{figure}[h!]
  \centering
  \begin{subfigure}[t]{0.49\textwidth}
    \includegraphics[width=\textwidth]{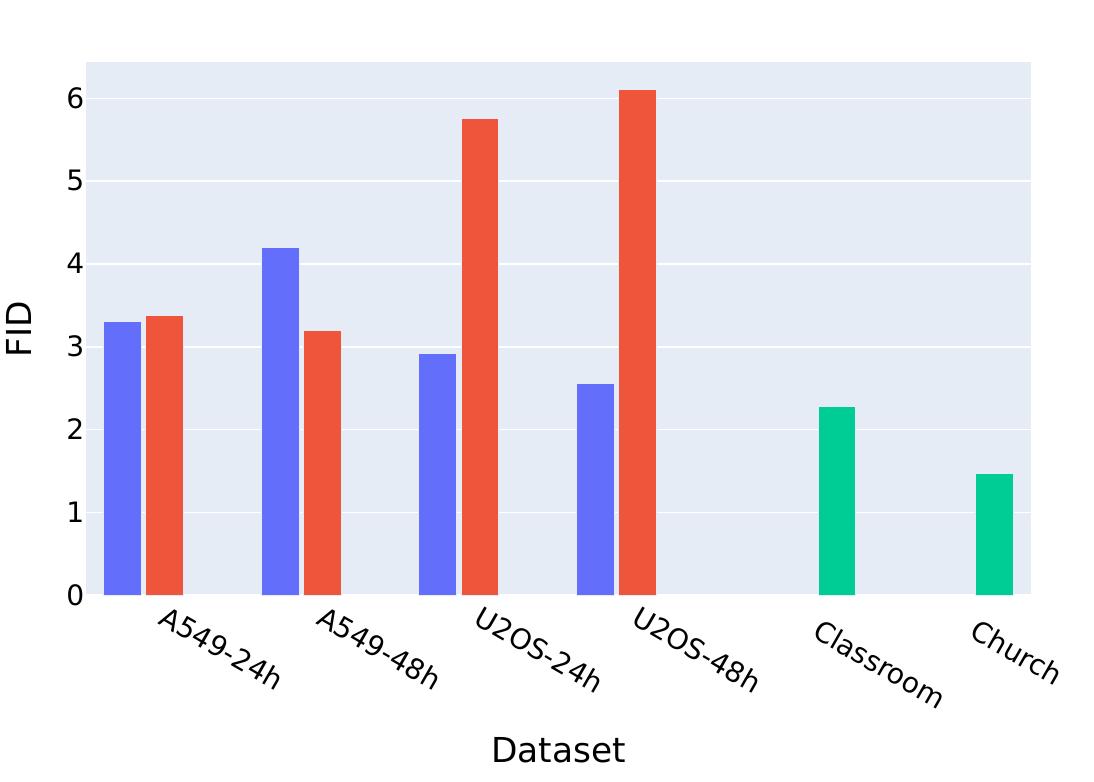}
  \end{subfigure}
  \hfill
  \begin{subfigure}[t]{0.49\textwidth}
    \includegraphics[width=\textwidth]{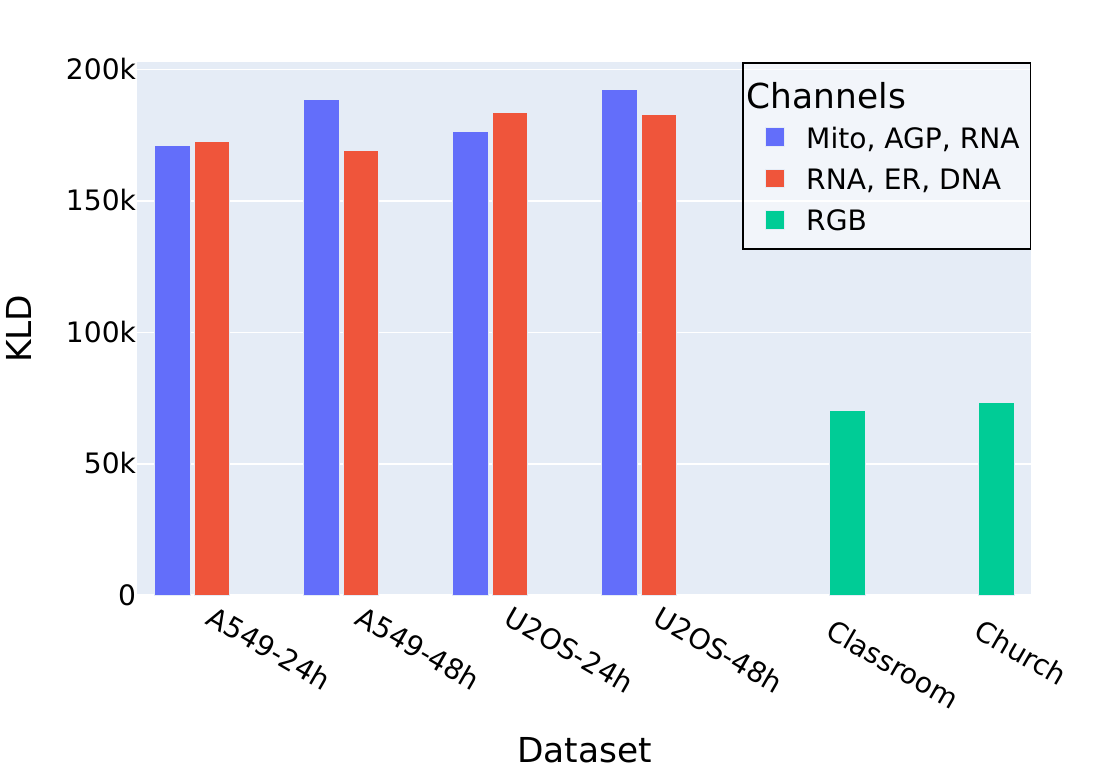}
  \end{subfigure}
  \vspace{-3pt}
  \caption{FID (left) and KLD (right) scores across the various data subsets, computed after a single run of SD-VAE applied to all images. FID scores are computed using real and reconstructed images. All images were featurized using InceptionV3. KLD scores are presented as the mean values computed across all samples within each dataset. Standard deviations are approximately 19k, 8k, and 10k for Cell Painting, Classroom, and Church images, respectively.}
  \label{fig:fid_kl_combined}
\end{figure}

In Figure \ref{fig:fid_kl_combined} we show instead the FID and KLD cross the LSUN and CPJUMP1 subsets, illustrating how the Cell Painting latents appear to be slightly less regularized than the latent embeddings of the natural images following application of the SD-VAE encoder.

Note that the results may be affected by how we chose to group the channels for the InceptionV3 model, which requires 3-channel inputs. This is particularly relevant given the noticeable variation in FID scores across channels. At the 24-hour time point, some channels show nearly twice the FID values compared to others (see Figure~\ref{fig:fid_kl_combined}). We leave examining the effects of channel grouping on the metrics to future work.

\begin{figure}[h!]
  \centering
    \includegraphics[width=0.49\textwidth]{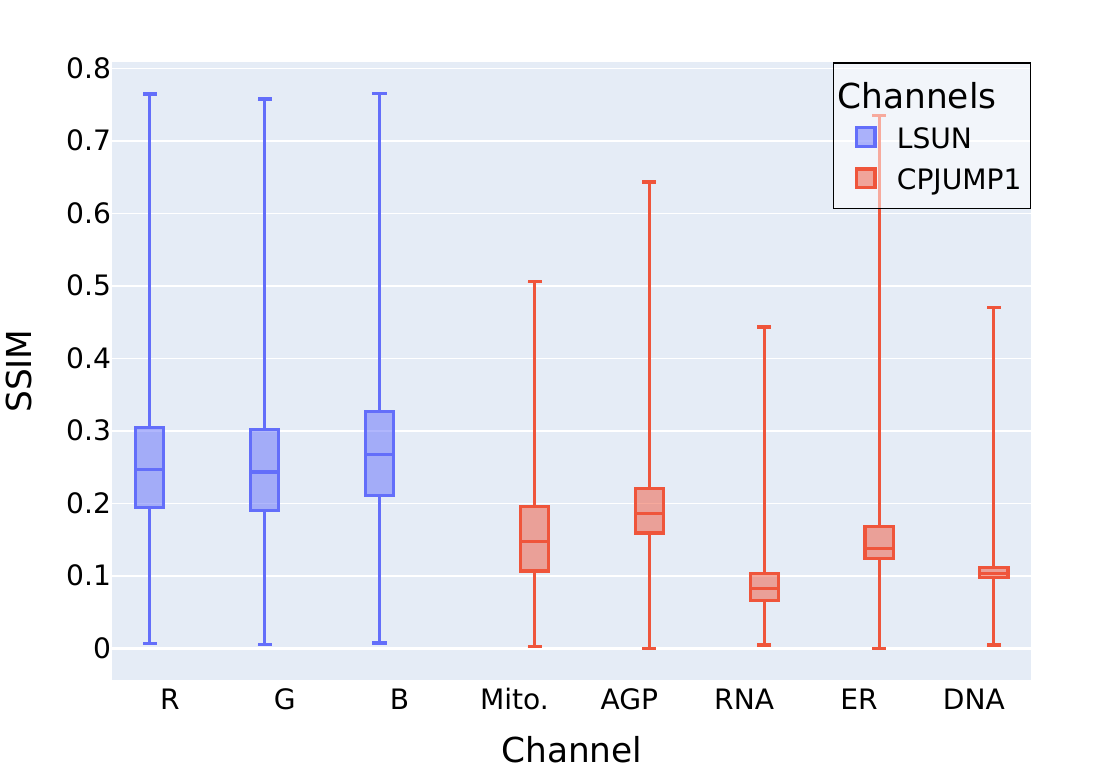}
    \caption{Box plot showing SSIM values across the LSUN and CPJUMP1 datasets and their respective channels, computed after a single run of SD-VAE applied to all images. The central line within each box shows the median, while the box boundaries represent the 1st and 3rd quartiles.}
    \label{fig:ssim}
  \hfill
\end{figure}

Interestingly, we observed that the OpenPhenom features benefit more from mean aggregation, while InceptionV3 features consistently perform better with median aggregation (Table \ref{table:aggregated_full}); this may reflect further differences in how the two models handle outliers or noise in morphological profiles. Regardless of the aggregation method, InceptionV3 features outperformed OpenPhenom accross all evaluated conditions.

\begin{table}[ht]
\centering
\caption{Fraction retrieved (FR) across cell types and time points using different aggregation strategies (mean plate scaling of DL features). All experiments using random features resulted in a FR of 0. The reported values are averaged over three independent runs of the complete pipeline (SD-VAE and information retrieval). For experiments using mean aggregation, the maximum standard deviation is $\sigma=0.006$, and for median aggregation $\sigma=0.008$. Experiments conducted on original images by design show no variation ($\sigma=0$). Best value for each cell line and time point is shown in bold.}
\medskip
\begin{tabular}{lll|cccc}
\toprule
 & & & \multicolumn{2}{c}{\textbf{A549}} 
& \multicolumn{2}{c}{\textbf{U2OS}} \\
\cmidrule(lr){4-5} \cmidrule(lr){6-7}
\textbf{Features} & \textbf{Data} & \textbf{Aggregation} & \textbf{24h} & \textbf{48h} & \textbf{24h} & \textbf{48h} \\
\midrule
OpenPhenom & Original      & Mean   & 0.804 & 0.925 & 0.774 & 0.719 \\
OpenPhenom & SD-VAE        & Mean   & 0.821 & 0.915 & 0.827 & 0.768 \\
InceptionV3 & Original     & Mean   & 0.846 & 0.935 & 0.833 & 0.768 \\
InceptionV3 & SD-VAE       & Mean   & 0.846 & 0.915 & \textbf{0.852} & 0.750 \\
\midrule
OpenPhenom & Original      & Median & 0.722 & 0.882 & 0.817 & 0.660 \\
OpenPhenom & SD-VAE        & Median & 0.729 & 0.879 & 0.836 & 0.697 \\
InceptionV3 & Original     & Median & 0.873 & \textbf{0.961} & 0.837 & \textbf{0.837} \\
InceptionV3 & SD-VAE       & Median & \textbf{0.906} & 0.951 & 0.847 & \textbf{0.837} \\
\midrule
CellProfiler \cite{Chandrasekaran2022ThreeMI} & Original & Median & 0.761 & 0.954 & 0.775 & 0.663 \\
\bottomrule
\end{tabular}
\label{table:aggregated_full}
\end{table}

We noticed that achieving high FR is more difficult for the U2OS cell line than for A549, even though U2OS was included in the OpenPhenom training set. The difference for this discrepancy remains unclear, but may be due to greater phenotypic heterogeneity in the U2OS cell line.

\subsection{Hardware and Compute Resources}

To facilitate parallelization across multiple GPUs, the datasets are divided into several subsets. To avoid the overhead of saving latent and reconstructed images, the MAE, SSIM, EMD, and KLD metrics are computed and saved in real time during inference in a batch setting. Features extracted from InceptionV3 and OpenPhenom are also saved. Inferencing the datasets was completed in just a few hours by leveraging dozens of NVIDIA A40 GPUs in parallel (up to 340 GPUs). On our facilities, running the whole pipeline, from downloading the images, preprocessing them, featurizing them, all the way to the post-processing and final analysis takes around 8 hours, if exploiting the parallelism of multiple jobs at the same time.

\end{document}